# Radiation belt electron precipitation in the upper ionosphere at middle latitudes before strong earthquakes

G. Anagnostopoulos,  V. Rigas and E. Vassiliadis

*Democritus University of Thrace*
*Space Research Laboratory*

**Abstract.** In this article we present examples of a wider study of space-time correlation of electron precipitation event of the Van Allen belts with the position and time of occurrence of strong (M>6.5) earthquakes. The study is based on the analysis of observations of electron bursts (EBs) with energies 70 – 2350 keV at middle geographic latitudes, which were detected by DEMETER satellite (at an altitude of ~700 km). The EBs show a relative peak-to-background increase usually < 100, they have a time duration ~0.5 – 3 min, energy spectrum with peaks moving in higher energies as the satellite moves towards the equator, and highest energy limit <~500 keV. The EBs are observed in the presence of VLF waves. The flux-time profile of the EBs varies in East Asia and Mediterranean Sea at the similar geographic latitudes, due to the differentiation of the magnitude of the earth's magnetic field. The most important result of our study is the characteristic temporal variation of electron precipitation variation which begins with incremental rate several days / a few weeks before major earthquakes, then presents a maximum, and finally become weaker, with a minimun or a pause of the electron precipitation some hours before the onset of the earthquake.

## 1. Introduction

**Electromagnetic methods of earthquake precursory signals research**

The last few decades significant effort has been made to detect and interpret electromagnetic phenomena related to seismic activity, and, due to this work, characteristic physical changes have been confirmed as precursor signals of preparing earthquakes (EQs). To this end, several methodologies have been developed, which use earth based instrumentation to detect electromagnetic variations occurring in the lithosphere (Kopytenko et al. 1990; Hayakawa et al. 1996a; Varotsos et al. 1998).

Other studies have shown that, before strong earthquakes, characteristic electromagnetic interactions occur in the ionosphere, which are observed as plasma variations or electromagnetic emissions, by either earth based (Hayakawa et al. 1996b; Pulinets & Boyarchuk 2004) or space based (Parrot 2006) instrumentation. For instance, Fig. 1, at left, shows such an investigation of the ionosphere with VLF broadcasting between a transmitter in Japan (JJU) and a receiver in Russia (PTK), and at right, some epicenters of earthquakes in Japan, studied by the above transmitter-receiver system (Rozhnoi et al., 2007).

In addition to the earthquake related electromagnetic phenomena taking place in the lithosphere and ionosphere, electromagnetic processes related to seismic activity affect the trapped population of the radiation (Van Allen) belts and several authors have reported satellite measurements suggesting radiation belt energetic particle flux variations during enhanced VLF electric field activity before earthquakes (Ginzburg et al. 1994; Galper et al. 1995; Aleksandrin et al. 2003; Pulinets & Boyarchuk, 2004 and references therein; Sgrigna et al. 2005). Research results based on DEMETER satellite measurements suggested no correlation between anomalous electron flow and seismic activity within 18 hours from the EQ occurrence (A. Buzzi, 2007), whereas a more recent study we performed suggested that energetic electron precipitation occurs several days before strong EQs and that this phenomenon follow a distinct temporal pattern, at least for EQs in Japan, which were examined statistically (Anagnostopoulos et al., 2010). Therefore, the question whether radiation electron anomalous bursts are EQ precursory signals seems that it has not been clearly answered. As a result, the phenomenon of energetic electron precipitation in the upper ionosphere has not been seriously considered as a valuable EQ prediction tool by recent times.

In this paper, we extend our previous analysis of EBs (Anagnostopoulos et al., 2010), and we present the main observational characteristics of EBs occurring several days to a couple of weeks before strong EQs, and we provide evidence that satellite observation of radiation electron precipitation in the upper ionosphere-plasmasphere is a hopeful tool in the EQ prediction research.

## 2. Theoretical framework: Earthquake induced radiation electron precipitation

### 2.1 Radiation belts (Van Allen).

Within the Van Allen belts, electrons and ions follow spiral trajectories around the field line of Earth's magnetic field, while drifting perpendicular to the magnetic field, in opposite direction protons and electrons; at the same time, ions and electrons moving towards higher latitudes are reflected in the strong magnetic field and are trapped within the inner and the outer Van Allen belts (Figures. 2 and 3).

Fig 4, in the left side, shows the average spatial distribution of electrons with energy > 1 MeV around Earth, in xz plane (SPENVIS). We distinguish the two radiation (Van Allen) belts, with the high electron intensities in red colours. In the center and the right side of Figure 4, two examples of DEMETER satellite (see below) semi-orbits are shown. The three curves indicate electron measurements from the IDP experiment (see below) onboard the DEMETER satellite in the energy ranges 72 - 526 KeV (Cnt1), 526 - 971 KeV (Cnt2), 971 - 2350 KeV (Cnt3). During both DEMETER's semi-orbits, two peaks in electron intensities are evident, in particular at low energies (red lines) in the left as well as in the right edge of each graph, which correspond to the edges of the two radiation belts in the north and the south hemisphere. The central panel shows electron measurements during quiet conditions (which corresponds to the average electron distribution shown in the left side of the figure). In the right panel, two electron bursts (EBs) are clearly seen (indicated by arrows), which are superimposed on the background intensity of the normal electron radiation belt profile (compare with the example in central panel), and are marked with bright red curve in the low energy band (see below). This kind of electron bursts is the subject of the present study.

### 2.2 Radiation belt electron precipitation caused by seismic activity

The radiation belt EBs associated with an earthquake at low and middle geographic latitudes manifest radiation electron precipitation and this process has been considered as one of the links in a long chain of electromagnetic processes in the lithosphere-atmosphere-ionosphere-plasmasphere-magnetosphere coupling. This process is caused by the anomalous electric field generating above the seismic preparation zone and the penetrating of in the ionosphere. The electric field causes electron density irregularities in the ionosphere and a subsequent increase in VLF wave activity in the magnetosphere with mechanisms still under investigation. It has been suggested that, radiation belt electron cyclotron resonance interaction with VLF waves is responsible for changes at the phase velocity of radiation belt electrons, which eventually steer electrons inside the loss cone (Inan et al. 1978; Pulinets & Boyarchuk, 2004). Therefore, a portion of the pre-trapped electrons travel to the Earth, and are observed by the satellite in the ionosphere at low altitudes from Earth's surface.

Furthermore, an electron burst accompanied by VLF waves in the north (south) hemisphere is also observed at the conjugate region in the south (north) hemisphere, due to the magnetic mirror effect (Fig. 6, left side). Our analysis of observations suggest that the conjugate event is usually weaker than the event above the seismic preparation zone (see below).

## 3. DEMETER mission and Instrumentation

In this paper we will present results from an elaboration of EBs observed by the French satellite DEMETER (Parrot, 2006). The satellite DEMETER (Fig. 7, left frame / http://demeter.cnrs-orleans.fr/) was designed and built specifically for the research of earthquake precursor signals (DEMETER: **D**etection of **E**lectro-**M**agnetic **E**missions **T**ransmitted from **E**arthquake **R**egions). The Demeter satellite was launched on June $29^{th}$, 2004, in a sun-synchronous almost polar (inclination $98.3^0$) orbit, at ~710 km altitude and performs 14 orbits per day, with an orbital period of ~ 102 min (right panel of Figure 7).

The primary scientific objective of DEMETER is the detection and characterization of ionospheric and radiation belt disturbances related to seismic or human activity. Consequently the scientific payload of the satellite is composed of several types of sensors measuring waves and particles, among which IDP and ICE: the **I**nstrument for the **D**etection of **P**articles and the Electric Field Instrument (**I**nstrument **C**hamp **E**lectrique). Data from these two instruments were used in the present study and are presented in Section 4.

The IDP electron spectrometer (Shauvaud et al. 2006) covers all energies from 0.07 – 2.5 MeV, in 256 energy bands. In this paper IDP electron intensities were averaged over three energy bands: 72 – 526 keV ("Cnt1"), 526 – 971 keV ("Cnt2") and 971 – 2350 keV ("Cnt3") (corresponding to red, green and blue lines, in corresponding figures). ICE's measurements are made over a wide frequency range of electromagnetic and / or electrostatic waves from DC to 3.175 MHz, subdivided in four frequency channels DC/ULF, ELF, VLF and HF. Here we use measurements from the VLF channel.

For the IDP instruments there are two modes of operation, the "burst" and the "survey" mode. The burst mode, with time resolution of 1 sec, is applied automatically when the satellite is overflying areas which have been identified as seismically active and are shown by red colour on the map of Figure 8 (bottom panel). When Demeter moves away from the seismically active regions, it returns to the "survey" mode, with a time resolution of 4 sec.

# 4. Observations

**4.1 Case studies**

We have examined energetic anomalous EBs detected by the IDP experiment along with VLF electric field activity from the ICE instrument onboard DEMETER for a period of ~4 years, from August 2004 to June 2008.

We chose to study mainly earthquakes in middle latitudes in East Asia (China and Japan) where the statistical sample of strong earthquakes is remarkable and the background intensity is relatively high (see below "Remarks over Greece"). In these areas EBs have certain common characteristics. Our investigation confirms the results of previous studies that EBs are observed before strong EQs during times of strong electric field VLF activity in large areas above the position of an EQ epicenter (Pulinets & Boyarchuk 2004, and references therein).

First we present an example of electron precipitation for the case of the deadly earthquake in China in May 2008. Figure 8 presents observations of the satellite DEMETER during the semi-orbit numbered 20609_1, in a relatively short distance from earthquake the epicenter ($103^0$E, $31^0$N), on 10 May, 2008, about 2 days before the earthquake occurrence. In Figure 8, we see from top to bottom, (a) the range of the electric field in the frequency range 0 - 18 kHz, (b) the intensity of electrons from the IDP experiment in energy regions 72 - 526 keV, 526 - 971 keV, 971 - 2350 keV, and (c) the semi-orbit 20609_1 projected on an earth's map. The global and local time, as well as the latitude and the longitude of the position of DEMETER during that time period have been noted below panel b. The small green circle marked in China indicates the epicenter coordinates.

From the middle panel, two anomalous EBs are obvious, indicated with vertical dotted lines in the figure and the thick segments of the DEMETER orbit in the bottom panel. The EB at right manifests electron precipitation near the epicenter of the EQ (northern hemisphere), and his conjugate, in the south hemisphere. The VLF electric field measurements in the top panel suggests the activity in the VLF electric field measurements was stronger near the epicenter of the earthquake than on the site of the conjugate EB in the southern hemisphere. Furthermore, VLF activity was observed in a long range of latitudes along the satellite almost polar orbit, starting from near EQ latitudes

Figure 9 was constructed with the same type as Figure 8, but for the semi-orbit 5864_1 of the day 8/8/2005, 8 days before the earthquake of August 16, 2005, in the eastern Honsu Japan. The typical increase in the intensity of electrons on 8/8/2005, with two EBs lasting ~ 3min, shows an example of the intense electron precipitation observed before the strong earthquake in eastern Honsu. Notice the VLF activity around the time of EB above the EQ epicenter and the weaker activity at conjugate region, in the southern hemisphere (as in the case of the earthquake in China on May 12, 2008, discussed above; Fig. 8).

Figure 10, at the top, shows a dynamic energy spectrum of energetic electrons from the DEMETER semi-orbit 05849_1 in August 8, 2005, passed near the epicenter of the earthquake of August 16, 2005, at eastern Honsu, Japan. The bottom panel of Figure 10 shows electron intensities at low, middle and high energy bands Cnt1, Cnt2 and Cnt 3. The comparison of the two plots suggests that the anomalous strong electron burst observed near the earthquake epicenter at ~11:23 UT shows a spectrum with a hardening toward the equator. The spectrum of this burst extends up to ~450 keV. The EB observed at the conjugate region is a little less intense and its spectrum extends up to energies of ~300 keV. In general, our investigation suggests that EBs

preceded EQ occurrence at middle latitudes show similar spectra, with a hardening toward the equator, most often extend up to 200 – 500 keV.

**4.2 Spatial-temporal distribution of energetic electron bursts before the earthquake M7.2 of August 16, 2005, in Japan.**

In Figure 13 we show the flux-time profiles of energetic electrons for all the night side orbits closest to the epicenter of the earthquake of August 16, 2005, at eastern Honsu, for each day from 4 to 16 August, 2005 (except for one orbit without data in August 11, 2005). It is evident that distinct EBs separated from the radiation electron belts structure appear from 5/8/2005 until 15/8/2005, with increasing peak-to background flux ratio p/b until almost the middle of the time interval examined, day 8/10/2005, and with decreasing values at the last days. Moreover, non EB was observed in the date of EQ, August 16, 2004. Table 2 gives full information for the longitude and latitude ranges covered by DEMETER for the time interval for each of the panels of Figure 13, and Figure 14 shows the segments of DEMETER orbit when the EBs seen in Figure 13 were observed. From the analysis of the form of each of the EB seen in Figure 13, we found that the p/b ratio ranged from 1.5 to 41, with a mean value at 8, and the EB time duration ranged from 50 sec to >3min (for only one case) with an average duration ~ 2 min, 23 seconds.

What is the most important result of this study is the characteristic pattern for the temporal variation of energetic electron precipitation before the occurrence of the EQs we studied. This has been inferred in case of Figure 13, as a special variation of the value p/b over the period from 5/8/2005 until 15/8/2005. Here we examined the possible influence of the electron precipitation during this period in broad regions over the globe. Figure 15b (bottom panel) shows the daily number $n_D$ of the Cnt1 low energy EBs (panels b and e in Figures 1 and 2) observed by DEMETER for the time interval 2-20 August, 2008 in the longitude range $107^o$ – $177^o$ (EQ epicenter was at $142^o$). It is evident a peak in $n_D$ 8 days before the EQ. About 15 days before the earthquake occurrence, $n_D$ starts increasing it shows a minimum on August 16, the Day of EQ. Panel a of Figure 3 shows the daily number $n_D$ of EBs observed by DEMETER all over the globe for the time interval 2-20 August, 2008. We see that the distribution pattern of $n_D$ all over the globe is similar to that observed in the area above the EQ epicenter. At this point we decided to examine whether the two curves were identical or not. For this reason we determinate the value of $X^2$ distribution with 19 degrees of freedom and we found from suitable tables that the two curves are identical at a 2.5% significant level. We understand the similarity of the two distributions of DEMETER daily number electron bursts as suggesting a strong dependence of the energetic electron flux all over the globe from the electron precipitation caused by the "local" electromagnetic processes of the EQ preparation region, due to fast electron drifting around the globe.

Figure 11, right side, shows the distribution of the EBs recorded by DEMETER as a function of longitude, where as latitude ϕ=0 has been defined the location of the of the earthquake's epicenter. Distributions are presented for three days: at the beginning of the phenomenon (03/08/2005), the day when most EBs were observed (08/08/2005) and the day of the earthquake occurrence. It is obvious that the number of EBs varies according to the shape of temporal evolution shown in on the left side of the figure. The important new information contained in this image is in the form of the spatial distribution of EBs on the day of earthquake: a quiet time period above the epicenter of the earthquake concerning electron precipitation!

### 4.3 Statistical analysis of large earthquakes in Japan

Study of a large number of strong (M>6) earthquakes at middle latitudes in China and Japan has led us to the conclusion that the temporal evaluation of energetic precipitation in the upper ionosphere follows the pattern of August 16,2005 earthquake that was shown in Figure 11. This pattern is more clear as much as other (previous or future) strong earthquakes are separated by longer times (days) and / or take place at more distant places. Our statistical analysis of large (> 6.7M) earthquakes in Japan for the time interval from August 2004 to June 2008, has shown that 7 out of 7 earthquakes, for which complete DEMETER data set were available, and the EQs were separated by more than 1 day, the temporal distribution of the daily number of local (+/- 35 deg.) EBs follows the MaxMin pattern of Figure 11. Table II provides appropriate information for these 7 earthquakes. The first 6 columns after numbering show the time, epicenter coordinates (longitude, latitude), magnitude and the depth of the EQ. In the last column, we give the time difference (in days) between the beginning time of the electron precipitation and the time of the earthquake occurrence time.

### 4.4 Observations over Greece.

In Table 1, on the left, we show data concerning an example of DEMETER data for a Greek earthquake. The data are given if the format used in previous example for EQs in China and Japan (Fig. 8 and 9). The energetic electron bursts detected above Andravida show a different behavior, with many short duration intensity peaks in a different (low) intensity background intensity, instead of one burst of longer duration (2-3 min) observed in China and Japan. In order to show the main cause of the flux – time profile differentiation in Greece and China / Japan we also show in Figure 14 in the right side, at the bottom panel, the average intensity of the ~ 200 keV energy electrons for one year (Oct. 2005 - Oct. 2006), and at the top panel the magnetic cell constant L, which varies with latitude.

Although Greece is approximately in the same geographic latitude as Japan, as indicated by the white horizontal line at ~ 39$^o$N in the map with the average electron intensity (bottom right panel), the electron precipitation is detected in different form in flux – time profile. This is due to the different magnetic field, which in general appears an asymmetry due to the South Atlantic Anomaly, which allows a greater amount of particles to reach the orbit of DEMETER in the broad area around the SAA. It also results in differentiation in the average background flux over the globe. In some areas, such as Greece and the region around the Mediterranean Sea, where the background intensity is low, the form of the flux-time profile of EBs preceded EQs is different than that in East Asia.

# 5. Discussion and Conclusions

The precipitation of energetic electrons from the Van Allen belts in the upper ionosphere, at middle and low latitudes, is generally a well known phenomenon in the scientific literature. The major sources of this process were considered powerful military transmitters and natural processes preceding strong earthquakes. However, so far has been accepted, without experimental statistical estimation, that the precipitation of electrons from the Van Allen belts in the upper ionosphere was due to manmade VLF transmitters. This is rather due to the relation of Communications with the Ionosphere that is an old and generally known scientific component of the modern civilization, whereas the relation of earthquakes with radiation belt electron precipitation concern a new and controversial research direction. In another paper in this issue we evaluated that the most powerful manmade transmitter NWC in western Australia contributes to a percentage as small as ~1% on the EBs at middle latitudes (Sidiropoulos et al); therefore, other (natural) processes are in general responsible for the ~70-500 keV electron precipitation at those latitudes.

The present study was based on results from a systematic study of the DEMETER satellite measurements from 2004 until 2010, indicating that the phenomenon of inner Van Allen belt electron precipitation at low altitudes is a permanent feature before very strong earthquakes (Anagnostopoulos et al. 2010), but also very often observed before EQs with magnitude M > 5 - 5.5.

Furthermore, our research has demonstrated that there is a permanent characteristic temporal variation in behavior of the flow of inner Van Allen radiation belt electrons in the upper ionosphere that lasts for several days to a few weeks before earthquake.

The phenomenon looks like a "rain" of electrons from the Van Allen belts, which begins with incremental rate several days to a few weeks before major earthquakes, then presents a maximum, and finally become weaker, with a minimun or a pause of the electron precipitation some hours before the onset of the earthquake (MaxMin pattern). The effect is observed locally, over the earthquake preparation zone, but in cases of strong earthquakes the precipitating electrons control the drifting electrons over wide ranges of longitudes.

From the time scale (of a few weeks to several days) that the EQ precursory process lasts, we understand the results of Buzzi (2007), who found no correlation between temporary electron flow and seismic activity, in a statistical study limited in a time window of (+ / - ) 18 hours around the earthquake occurrence time.

This kind of temporal evolution we found in space measurements of energetic electron precipitation, both the time scale and the MaxMin pattern, are known for several electromagnetic geophysical, ionospheric and space physics parameters before strong EQs (Pulinets & Boyarchuk 2004; Ouzounov et al. 2006). For instance, we remind the characteristic temporal pattern of the ULF magnetic field activity detected by ground magnetometers some weeks before an earthquake (Hayakawa et al. 1996b). In this case, the magnetic field ULF polarization ratio Z/H starts increasing gradually, it reaches a maximum value, and then, around the time of the EQ shows a deep minimum. It has been supposed by Hayakawa et al. (1996a) that these ULF emissions are related to earthquake preparation processes. Some other physical parameters also show a similar pattern with a peak before the EQ occurrence (Ouzounov et al. 2006). The similarity of the results of the present work on the temporal flux profile of the precipitating radiation belt energetic electrons with the pattern of the magnetic field polarization ratio Z/H reported by Hayakawa et al. (1996b) by comparing the two time profiles in Fig. 14.

We also found that the electron precipitation that is observed as bursts in electron intensity have different characteristics above Greece than in case of Japan. The phenomenological differentiation of the observational characteristics of the phenomenon and, consequently, some aspects of the generating process of the EBs varies with the coordinates of the EQ epicenter, due to the differentiation of the Earth's magnetic field over the globe.

We found that this pattern, with a Maximum flux several days before and a Minimum flux around the earthquake occurrence time, clear in 7 of the 7 >6.7 EQs occurred in or near Japan (long. 135 – 155 $^o$E) between August 2004 – June 2008.

Our investigation suggests that the Max-Min e-flux pattern appears more clear when we study the occurrence of a single EQ, that is when no other large EQs occur in close areas / times.

Furthermore, it is important to note that when we study the behavior of a single EQ, the Max-Min electron-flux pattern is not only clear above the area of the EQ epicenter, but it appears almost the same when averaging the daily number $n_D$ of the EBs observed by DEMETER over most of the globe (Fig. 12). This finding suggests that most probably the earthquake preparation process is related with strong enough electromagnetic phenomena in the atmosphere-ionosphere-radiation belts that they can control the global energetic electron precipitation at low altitudes.

We understand the radiation belt energetic electron precipitation related with an EQ as a ring in the chain of electromagnetic processes starting above the earth's surface with the anomalous electric field causes an increase in the VLF electric field wave activity in the magnetosphere. It has also been suggested that a cyclotron resonance interaction between VLF waves and radiation belt energetic electrons is responsible for variations in the electron velocity phase space (Inan et al., 1978). As a result, a part of the radiation belt trapped electrons change their pitch angles and can travel toward the earth and they are observed in the inner boundary of the magnetosphere and the ionosphere (Inan et al. 1978; Pulinets and Boyarchuk 2004). It is important to note that the ionization produced by the precipitation of electrons from the radiation belts in the lower ionosphere, produce disturbance in the D region of the ionosphere at night, which is then observed both from ground and space instrumentation (Pulinets & Boyarchuk, 2004). A diagram of the processes in the system atmosphere - ionosphere – inner magnetosphere which are induced by seismic activity is shown in Figure 16.

The results of the present study may open a new window in the earthquake prediction research. Statistical studies on a classification of the features of the radiation belt energetic electron events dependent on the spatial / spectral structure of the radiation belt above the EQ epicenter, the determination of the EQ epicenter magnitude, are in progress, based on existed preliminary results. We emphasize that the methodology of radiation belt energetic particle events as earthquake precursor signals has the particular advantage of studying distant earthquake preparation regions, because of the fast eastward drifting of the energetic electrons.

Generally speaking, we express the hope that an international strengthening of cooperation of various research groups working on different methodologies of electromagnetic EQ precursor signals, can achieve a high confidence level in earthquake prediction, at least in some cases, in the future years. Figure 17 presents, from left to right, a sketch of a system of synergy between earth and space based observations of the ionosphere and atmosphere, space observations of plasmasphere / radiations belts and earth based observations of geophysical and electromagnetic parameters. |The operation of data and risk analysis centers of such combined measurements is considered to be the next step in earthquake prediction research.

**Acknowledgments.** The authors thank Dr Parrot and Pr. Sauvaud for providing the Demeter data.

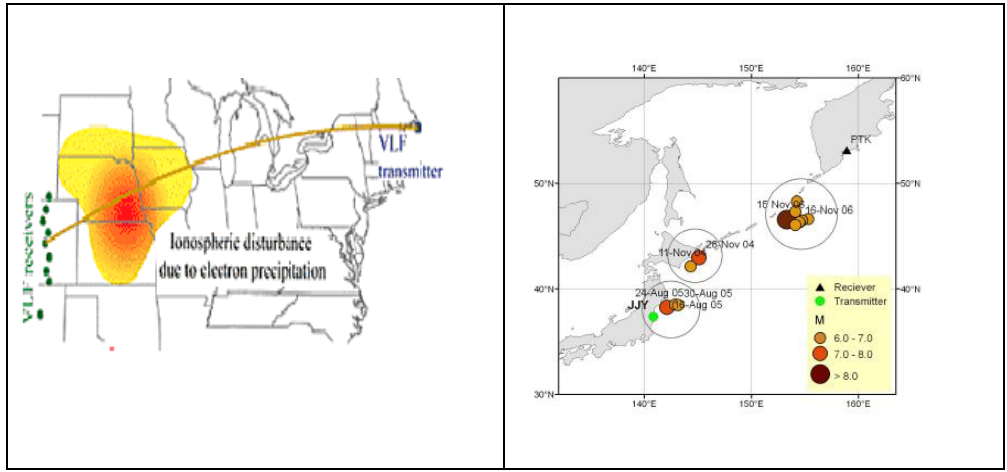

Figure 1. (Left) A transmitter-receiver system for ionospheric disturbances research in Japan. (Right) Epicenters of some large earthquakes, studied with this system by Rozhnoi et al. (2007).

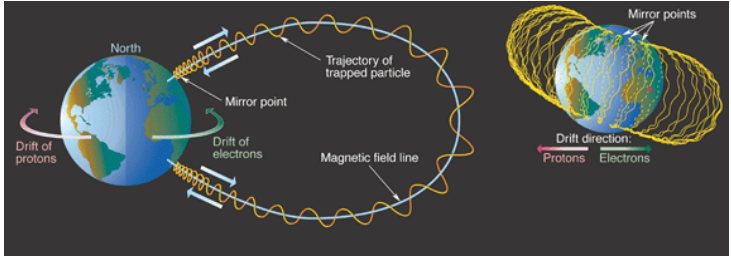

Figure 2. The motion of charged particles trapped within the Van Allen belts.

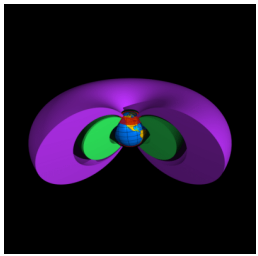

Figure 3. Schematic representation of the internal (green) and external (purple) belt.

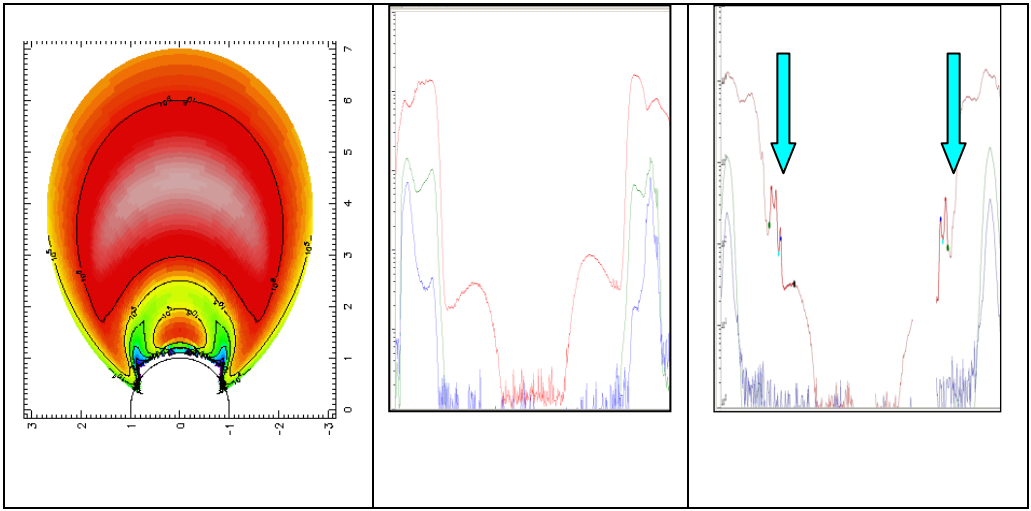

Figure 4. (Left) Distribution of (>1 MeV) electron density at the xz plane (SPENVIS). (Center) Example of electron intensity measurements at low (red), medium (green) and high (blue) energies in the whole region from 0.07 to 2.5 MeV, for a semi-orbit during a normal (quiet) time period. (Right) Example of a semi-orbit with an electron precipitation event and its conjugate one the other hemisphere (north / south).

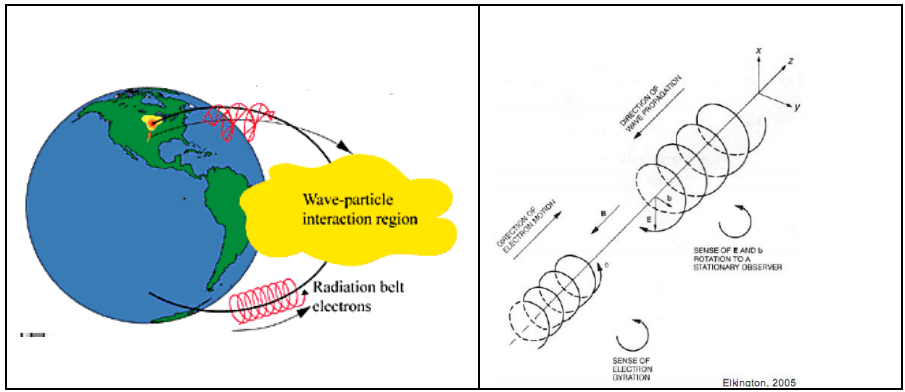

Figure 5. Cyclotron resonance interaction between VLF waves and radiation belt electrons.

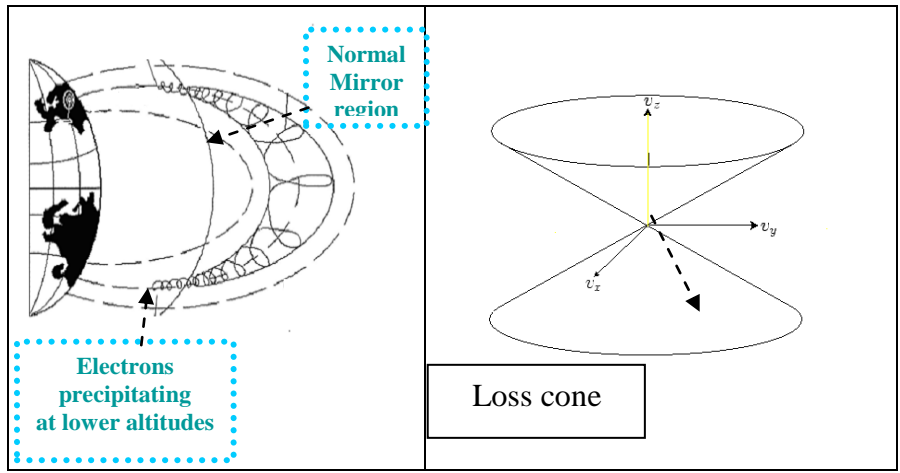

Figure 6. (Left) The cyclotron resonance interaction between VLF waves and radiation belt electrons causes changes in the electron phase velocity and steer them in the loss cone. Then, electrons travel deeper in the radiation belt and precipitate in the ionosphere-atmosphere. (Right) Particle with velocity direction within the loss cone do not "feel" a magnetic mirror and continue their motion toward stronger magnetic field.

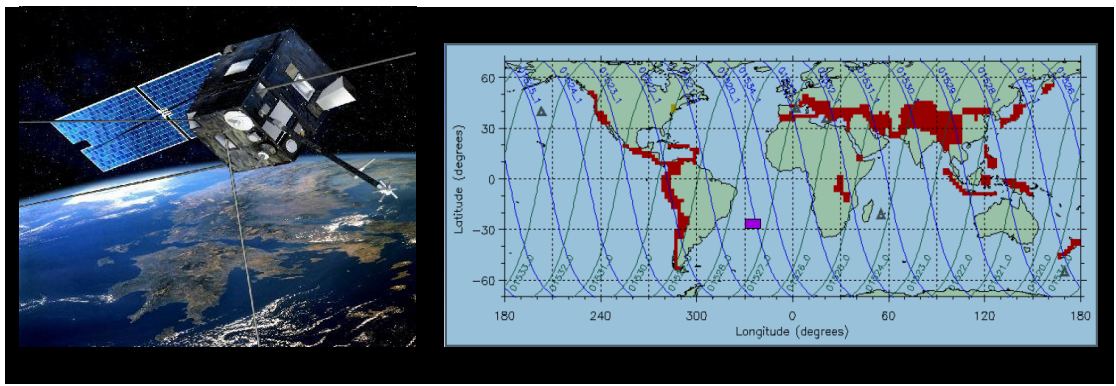

Figure 7. (Left) Scheme of the DEMETER satellite. (Right) Example of DEMETER satellite semi-orbits covered the Earth within a day. The red areas on the map indicate the most seismically active regions on earth.

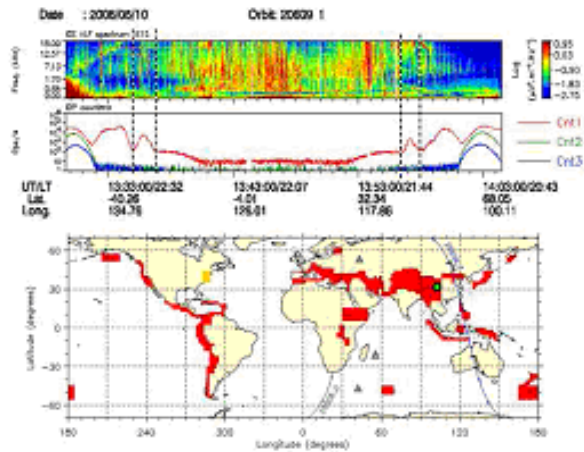

Figure 8. Electric field spectrum from the ICE experiment, electron fluxes in three electron bands from the IDP instrument and the corresponding trajectory (blue line) of the DEMETER satellite along with the position of the epicenter (green circle) of the earthquake in China (c, f) on May 12, 2008, for a trajectory two days earlier. The energetic electron "signal" with increased VLF activity as well as the conjugate one are enclosed by vertical dotted lines in two top panels (details in the text).

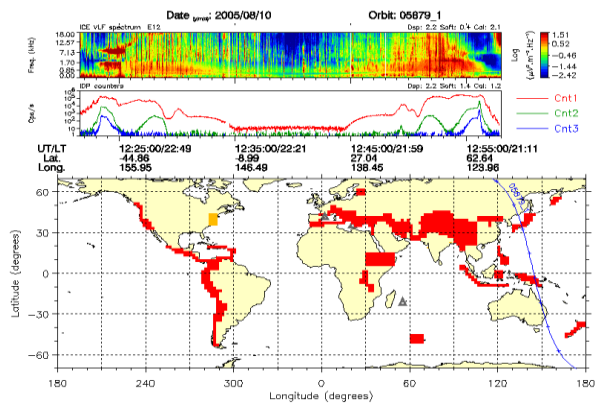

Figure 9. The same format as in Fig. 8, but for measurements concerning the earthquake occurred on August 16 near Japan, for a trajectory eight days earlier.

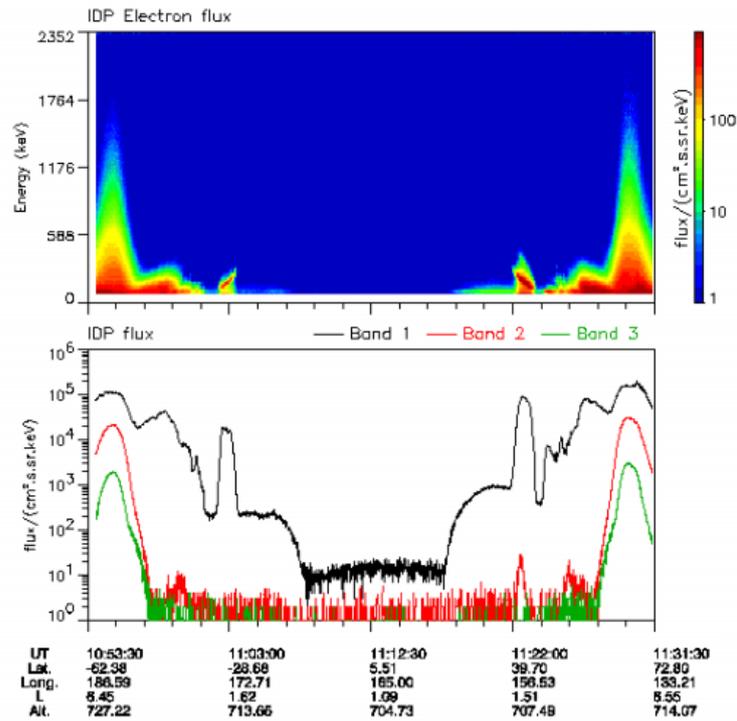

Figure 10. (top) Dynamic energy spectrum of energetic electrons from the DEMETER semi-orbit 05849_1 (August 8, 2005) passed near the epicenter of the earthquake of August 16, 2005, at eastern Honsu. (bottom) Electron intensities for three energy bands. (see in the text) The comparison of the two plots suggests that the anomalous strong electron burst observed near the earthquake epicenter show a spectrum with a hardening toward the equator.

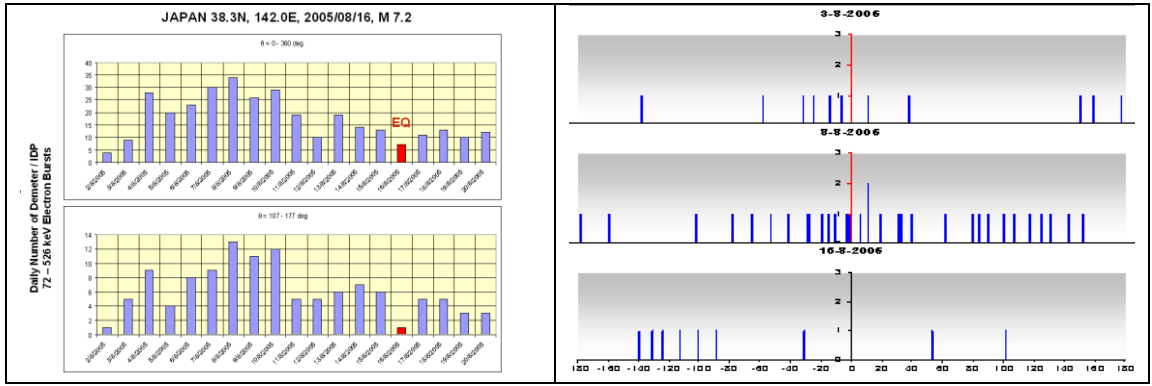

Figure 11 (left) Temporal distribution of the daily) number of the electron bursts across the globe (top) and related with electromagnetic processes above the preparation region of August 16, 2005 earthquake in Japan (bottom) during the period 2-20/8/2005. The correlation of the distributions is very important (Anagnostopoulos et al., 2010).

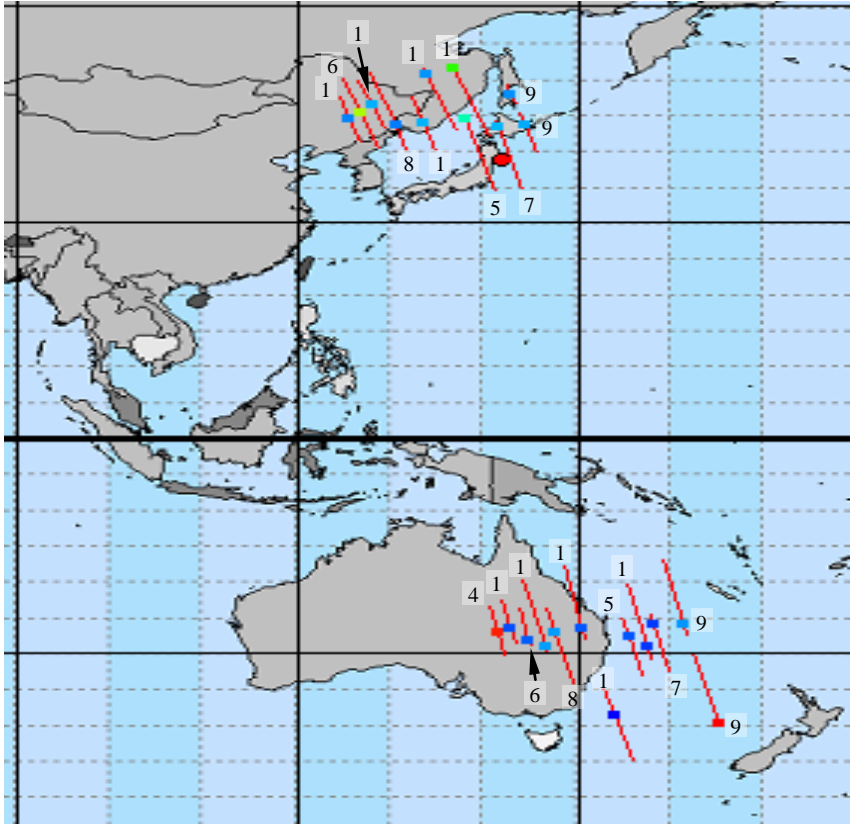

**Figure 12** Segments of DEMETER orbits for which electron bursts were observed for the time examined before the earthquake in Japan of August 16, 2005.

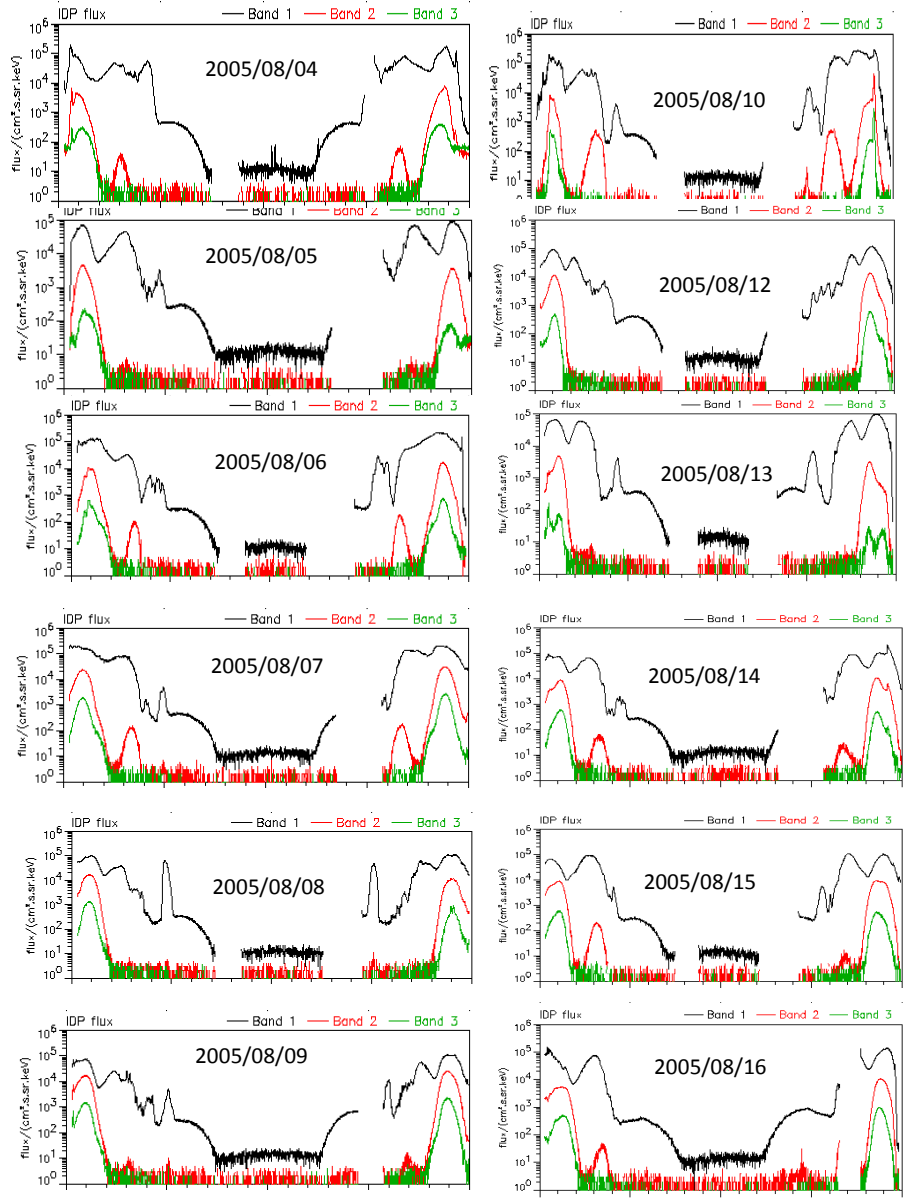

**Figure 13.** Flux – time profiles of energetic electrons above the epicentre of earthquake occurred in Japan in August 16,2005

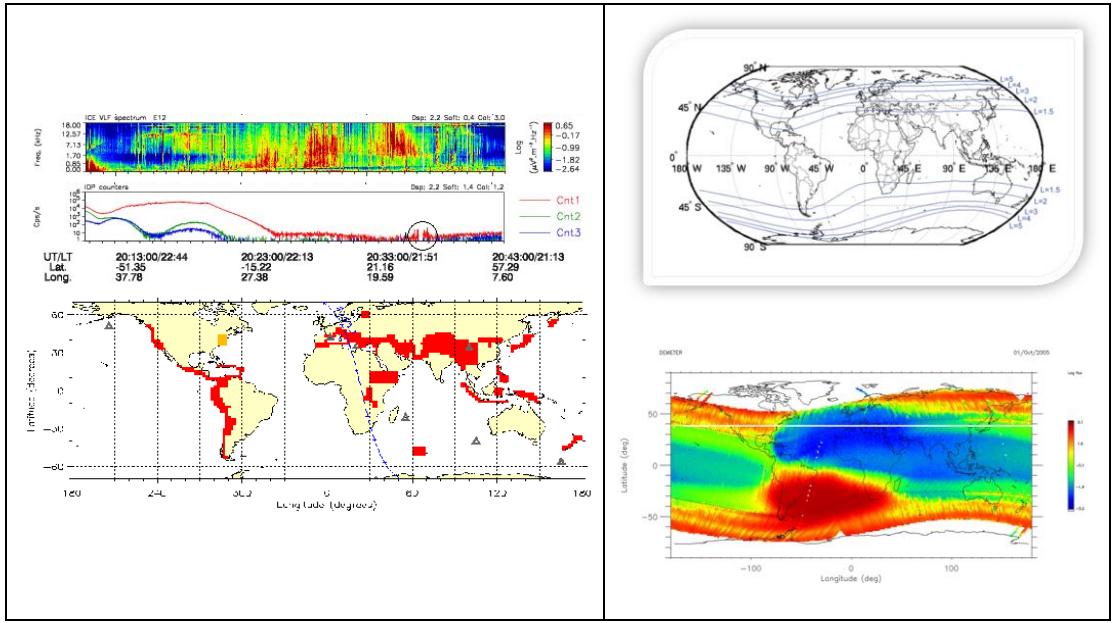

Figure 14. (Left) Example of Greek earthquake. (Right) H average intensity of the electron energy ~ 200 keV measured by one year (Oct. 2005 - Oct. 2006). It is obvious variation in the average intensity with longitude and latitude of each site due to the inhomogeneous magnetic field (Sauvaud et al., 2008)

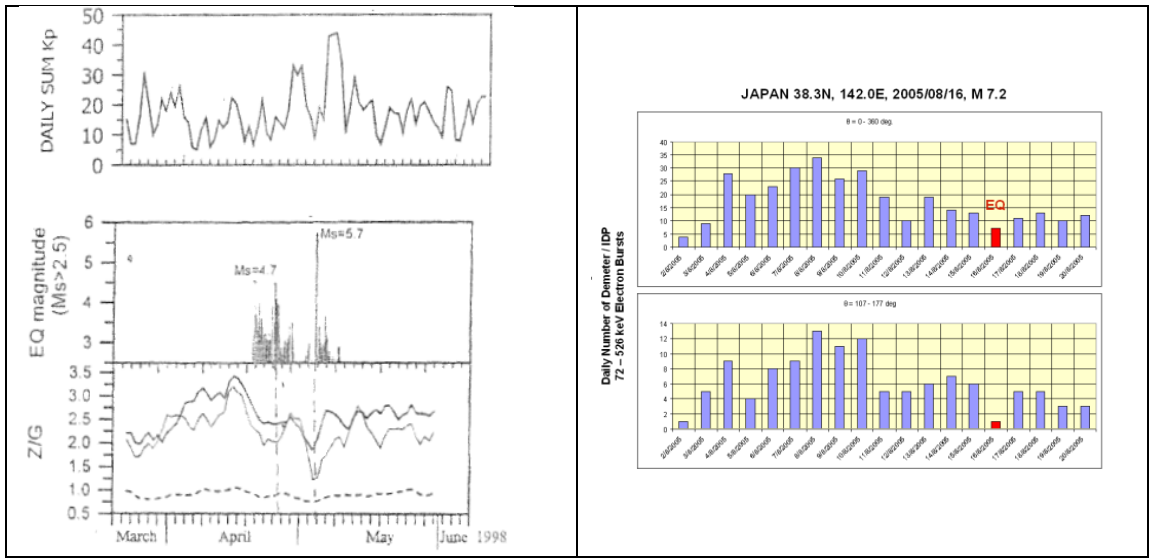

Figure 15 Comparison of temporal changes in terrestrial and space-based measurements. The two compared parameters show the same pattern.

Figure 16. Diagram of seismic activity induced.processes taking place in the system atmosphere - ionosphere - inner magnetosphere,

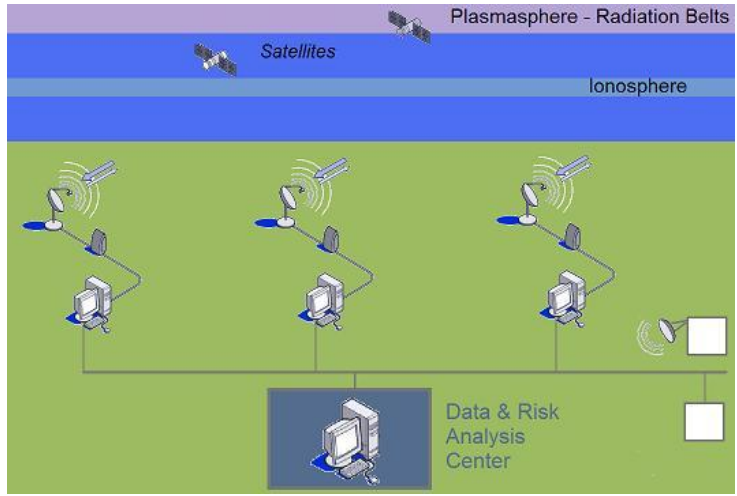

**Figure 17.** The operation of data and risk analysis centers of earth and space based observations is considered to be the next step in earthquake prediction research.

1
2

| orbit | s | Date | | | Start | | | | | | end | | | duration | |
|---|---|---|---|---|---|---|---|---|---|---|---|---|---|---|---|
| | | | | | Time UT | | | Lat | Long | Time UT | | | Lat | Long | m | s |
| 5792 | 1 | 2005 | 08 | 04 | 12 | 59 | 12 | -55.04 | 151.07 | 13 | 35 | 23 | 73.69 | 100.40 | 36 | 11 |
| 5806 | 1 | 2005 | 08 | 05 | 12 | 02 | 42 | -58.13 | 166.70 | 12 | 39 | 25 | 72.60 | 116.55 | 36 | 43 |
| 5821 | 1 | 2005 | 08 | 06 | 12 | 46 | 42 | -55.55 | 154.42 | 13 | 22 | 45 | 72.82 | 105.28 | 36 | 03 |
| 5835 | 1 | 2005 | 08 | 07 | 11 | 50 | 12 | -58.62 | 170.10 | 12 | 27 | 25 | 73.73 | 117.32 | 37 | 13 |
| 5850 | 1 | 2005 | 08 | 08 | 12 | 34 | 12 | -56.04 | 157.78 | 13 | 10 | 17 | 72.48 | 109.04 | 36 | 05 |
| 5864 | 1 | 2005 | 08 | 09 | 11 | 37 | 42 | -59.11 | 173.50 | 12 | 14 | 55 | 73.29 | 121.37 | 37 | 13 |
| 5879 | 1 | 2005 | 08 | 10 | 12 | 21 | 42 | -56.53 | 161.14 | 12 | 57 | 47 | 72.03 | 112.96 | 36 | 05 |
| 5908 | 1 | 2005 | 08 | 12 | 12 | 09 | 12 | -57.02 | 164.51 | 12 | 45 | 49 | 73.29 | 113.63 | 36 | 37 |
| 5923 | 1 | 2005 | 08 | 13 | 12 | 53 | 12 | -54.43 | 152.30 | 13 | 28 | 39 | 71.91 | 105.44 | 35 | 27 |
| 5937 | 1 | 2005 | 08 | 14 | 11 | 56 | 42 | -57.51 | 167.89 | 12 | 33 | 25 | 73.16 | 117.00 | 36 | 43 |
| 5952 | 1 | 2005 | 08 | 15 | 12 | 40 | 42 | -54.92 | 155.64 | 13 | 16 | 45 | 73.39 | 105.68 | 36 | 03 |
| 5965 | 1 | 2005 | 08 | 16 | 10 | 03 | 12 | -65.32 | 201.49 | 10 | 41 | 41 | 71.59 | 147.71 | 38 | 29 |

3
4
5 **Table 1**
6
7
8

| No | Date | Time (UTC) | Lat (deg.) | Long. (deg.) | M | Depth (km) | Duration (days) |
|---|---|---|---|---|---|---|---|
| 1 | 2005/08/16 | 02:46:28 | 38.28 N | 142.04 E | **7.2** | 36 | **15** |
| 2 | 2005/11/14 | 21:38:51 | 38.11 N | 144.90 E | **7.0** | 11 | **4** |
| 3 | 2006/11/15 | 11:14:13 | 46.59 N | 153.27 E | **8.3** | 10 | **7-8** |
| 4 | 2007/01/13 | 04:23:21 | 46.24 N | 154.52 E | **8.1** | 10 | **6** |
| 5 | 2007/03/25 | 00:41:57 | 37.34 N | 136.59 E | **6.7** | 8 | **12(4?)** |
| 6 | 2007/07/16 | 01:13:22 | 37.53 N | 138.45 E | **6.6** | 12 | **5-6** |
| 7 | 2008/05/07 | 16:45:18 | 36.16 N | 141.53 E | **6.9** | 27 | **12(17?)** |

9
10 Table 2. Statistical results for strong earthquakes in Jaapan (2004-2008)
11